# Harmonic Magneto-dielectric Study in Doped Perovskites, and Double Perovskite

Pooja Sahlot[1], Suchita Pandey[2], Adityanarayan Pandey[3], and A.M. Awasthi[1]*

[1]*UGC-DAE Consortium for Scientific Research, University Campus, Khandwa Road, Indore- 452 001, India*
[2]*Forensic Science Department, Police Training College, Indore- 452 001, India*
[3]*Department of Metallurgical Engineering and Materials Science, Indian Institute of Technology Bombay, Mumbai- 400 076, India*

*E-mail: amawasthi@csr.res.in

**Abstract**: Fundamental and harmonic magneto-dielectricity studied for varied perovskite systems-- $Pb_{0.98}Gd_{0.02}(Mg_{1/3}Nb_{2/3})_{0.995}O_3$ (A-site co-doped PGMN magneto-relaxor), $La_{0.95}Ca_{0.05}CoO_3$ (A-site doped spin-state LCCO), and $La_2NiMnO_6$ (double-perovskite LNMO multiglass) characterize intricately polarized phases. First-harmonic signal ($\varepsilon_2'$) of magnetically co-doped PGMN manifests finite polarization $P(H)$ below 270K, corroborated by the measured remnant $P$-$E$ traces. Second-harmonic ($\varepsilon_3'$) reveals the effect of random ***E***-fields causing electrical vitreousity, latter indicated by the divergent timescale of the fundamental response. LCCO features mixed-dipoles phase over appreciable temperature window, affiliated to the coexistent low-spins (LS) and intermediate-spins (IS). Across the 65K-start of IS-to-LS state transition (SST), dc- and ac-conductivities of LCCO exhibit mechanism-changeovers whereas the harmonic susceptibilities evidence IS/LS-interfacial hyper-polarizations. Below the 30K-end of SST, harmonics corroborate the vitreous phase of dipoles in the LS-matrix state. In the LNMO, positive and negative (dual) magneto-dielectricity observed is respectively attributed to the charge-hopping between $Ni^{2+}$ and $Mn^{4+}$ ions and the interfacial polarization. Second-harmonic signal here also features dispersion corresponding to the activation energy required for the electron transfer between Ni- and Mn-cations. Results from three different perovskite systems signify the combined importance of first- and second-harmonics, for a detailed understanding of electrical configurations.

## Introduction

Perovskites ($ABO_3$) have been studied extensively because of their broad application field. Some of their properties like the stability of structure, environment-friendly ceramics, and different electronic states with respective conductivity features provide good candidate for electro-catalyst, with wide use in rechargeable batteries (with high capacitance) and fuel cells [1,2]. Pb(B′B″)$O_3$ type mixed-perovskite relaxors have been widely studied because of their wide application in the field of multilayer ceramic capacitors, ultrasonic transducers for medical diagnosis, and many others [3]. $Pb(Mg_{1/3}Nb_{2/3})O_3$ is a well-known low-cost relaxor ferroelectric with large piezoelectric constant and electro-mechanical coefficient, high resistivity, high dielectric constant, and wide bandwidth response. These properties make it an excellent candidate for practical applications. PMN has cubic phase (Pm-3m) of paraelectric nature, with coexistent polar nano-regions (PNRs) of local rhombohedral (R3m) symmetry and chemically-ordered regions (Fm3m symmetry) of Mg/Nb superstructure [4,5]. Correlated dipole moments exist below 600K and freeze to PNRs below 220K [6]. This system has been vastly explored with A- and B-site substitutions, where the former is found to result in increased relaxor-like character while the latter in enhanced ferroelectricity [7,8]. Study of Gd-substituting Pb-site in PMN ceramics, with A/B-site vacancy, has explored multiferroic functionality of the system [9,10]. System shows glassy dynamics for dipolar structures present in the system, featuring enhanced relaxor-like character with increase in the Gd-doping. The system remains to be studied under magnetic field, to explore magneto-electricity in the system for prospective applications. For the present study, we have selected a 2% Gd-doped PMN sample having B-site vacancy, to avoid impurity phases introduced at high doping. Dielectric properties of the system are explored without and under 9T magnetic field. Further, non-linear dielectric response is studied and analyzed, which greatly details the dynamics and polarizability of the polar-regions under zero and 9T magnetic field.

B-site magnetic perovskite viz., $LaCoO_3$-based compounds have been said to be strong contenders against the platinum catalyst (oxidation-reduction) [11,12]. Dielectric spectroscopy is a nondestructive tool, where under the application of ac-electric field, impedance is measured. Real and imaginary parts of complex permittivity characterize respectively the energy stored in the structure of the system (dipolar response) and the dielectric losses (conductive response). $LaCoO_3$ has been reported to have multiple electrical and magnetic degrees of freedom accompanying the mixed (low/intermediate/high) spin-phases over appreciable temperature windows. $Co^{3+}$ exceptionally possesses high-spin (HS) state at high temperatures down to 90K, with antiferromagnetic correlations [13]. Proximity of crystal-field split in the octahedral environment and exchange energy for $Co^{3+}$ ions ($3d^6$ configuration) in the system trigger spin state transitions (SST) from HS state ($t_{2g}^4 e_g^2$ ) to intermediate spin (IS) ($t_{2g}^5 e_g^1$) below room temperature [14,15]. Upon further cooling, crystal field split energy value comes close to the exchange energy, stabilizing the low-spin (LS) state with $t_{2g}^6$ configuration. HS and IS states accompany Jahn-Teller monoclinic distortions in the parent rhombohedral structure [16]. Across the IS-LS spin state transition, study of the stretching modes via Raman spectra, corresponding to different spin states, confirmed the gradual change from one spin state to another, and their coexistence over a temperature window corresponds to an order of magnitude dispersive change in the dielectric properties [16]. This vast change is traced to a mixed electrical state, comprising of phase-coexistent dipoles affiliated to the respective spin-states [17]. Thus, the system qualifies for the dielectric study of the bulk electrical-interface, as the inter-domain region between two simpler electrical mediums is well-acknowledged for device applications, because it co-hosts the properties of both the media [18]. Analogous to the magneto-electricity in the composites, bulk-interfaces with no centrosymmetry in structurally single-phase materials can feature properties that are absent in the respective pure bulk media. Interestingly, La-site doping has evidenced intrinsic magnetic phase separation (magnetic glass, ferromagnetic clusters), which mandates further exploration [19]. In this regard 5% Ca-doped $La_{0.95}Ca_{0.05}CoO_3$ (LCCO) is chosen for the present study. This system has been studied in literature with signatures of SST in the dielectric properties (with giant magneto-dielectricity of up to 80%); featuring a magneto-thermally activated SST to Vogel-Fulcher/vitreous behavior of dipolar glass below ~30K [17]. Present paper studies the intermediate- to low-spin state crossover in LCCO via harmonic dielectric properties.

Further extending the study on perovskite-related structures, in double perovskites with general structural formula $A_2BB'O_6$, ordering of the transition metal (TM) ions yields anomalous physical properties, compared to the simple perovskite structures [20]. Charge ordering of these TM ions (+2 and +4 oxidation states) promotes charge localization of electrons; i.e. introduction of more insulating state compared to the parent perovskite [21]. $La_2NiMnO_6$ double perovskite, with high temperature ferromagnetic ordering and insulating character, has been proposed as a good candidate for the study of magneto-dielectricity, with applications as magneto-dielectric capacitors [22]. In $La_2NiMnO_6$, super-exchange ferromagnetic interactions among the ordered $Ni^{2+}$ and $Mn^{4+}$ ions results in near room temperature ferromagnetic ($T_c$ ~280K) semiconductor. Antisite disorder results either in the formation of anti-phase boundaries (discerning the ferromagnetic clusters) or the glassy magnetic phase [23]. Under the application of external magnetic field, alignment of the antiparallel ferromagnetic domains along the field direction has been reported to yield magneto-resistance (MR) of -6% in the system [24]. System has been reported to show Arrhenic dielectric relaxations with different magneto-dielectricity (MD) of -16% and +4% in different reports [25,26]. In one report, magneto-dielectric effect has been attributed to the combined effect of Maxwell-Wagner (MW) interfacial polarization and magneto-resistance, with adequacy for device applications [26]. In the other one, hopping of electrons among the TM sites is said to cause the magneto-dielectricity [25].

To characterize the dielectric state non-linear properties viz., the first-order harmonic ($\varepsilon_2' \equiv \chi_2'$) and second-order harmonic ($\varepsilon_3' \equiv \chi_3'$) are good tools of study. First harmonic signal directly correlates with the polarization state of the system, expressing itself as [27];

$$\chi_2' = -3\varepsilon_0^2 \mathrm{B} P \chi_1'^3 \quad (1)$$

Where, B is a smooth function of temperature expressed as, $4.5\times10^{-15}[T(\mathrm{K})\text{-}98.15]$ [28]. In the free-energy density functional, B is the proportionality factor for biquadratic polarization-term (~$P^2$). First harmonic is an indirect measure of the change in local polarization (as expressed previously), and second harmonic expresses itself in terms of the polarization and fundamental dielectric susceptibility as [27];

$$\chi_3' = -(1 - 18\varepsilon_0 \mathrm{B} P^2 \chi')\varepsilon_0^3 B \chi'^4 \quad (2)$$

Furthermore, since the dielectric properties are well coupled to the magnetic spin state, magneto-dielectric and magneto-harmonic $\left(\mathrm{MH}_{1,2} = \left\{\frac{\chi_{2,3}'(H)}{\chi_{2,3}'(0)} - 1\right\}\times100\right)$ characterizations of the perovskites are carried out to divulge novel insights to the intricate electrical phases.

## Experimental Techniques

$Pb(Mg_{1/3}Nb_{2/3})O_3$ and $Pb_{0.98}Gd_{0.02}(Mg_{1/3}Nb_{2/3})_{0.995}O_3$ ceramics are synthesized by columbite precursor method and characterized by X-ray diffraction, as reported in reference [10]. $La_{0.95}Ca_{0.05}CoO_3$ and $La_2NiMnO_6$ ceramics have been prepared using solid-state technique and XRD characterization of the sample is done by Bruker D8 Advanced Diffractometer using Cu-K$_\alpha$ radiation. From XRD pattern fit we have confirmed the single phase purity of $La_{0.95}Ca_{0.05}CoO_3$ specimen **(in supplementary file S1)**. Using precursors namely $La_2O_3$, NiO and $MnO_2$ in stoichiometric ratio $La_2NiMnO_6$ ceramic is prepared with final sintering at 1200°C. XRD **(in supplementary file S1)** and Raman measurements confirms formation of ordered double perovskite $La_2NiMnO_6$ structure. XRD pattern also depicts 4% secondary phase of NiO (antiferromagnet in cubic symmetry with centrosymmetric space group Fm3m) in the sample. Zero-field cooled (ZFC) and field cooled (FC) magnetization measurements with temperature and magnetic field were carried out using a Quantum Design SQUID-VSM. Raman scattering measurements were performed using HR800 Jobin-Yvon spectrometer with He-Ne laser of wavelength 632.8 nm. Low temperature fundamental and non-linear (first- and second-harmonic) permittivity measurements were performed in parallel plate capacitor cell in 1V-ac applied *E*-field, with NovoControl Alpha-A Broadband Impedance Analyzer, utilizing Oxford Nano-Systems Integra magnet-cryostat. Electric field induced *P-E* loop measurements are performed at 50 Hz with Precision Workstation of Radiant Technology, USA.

## Results and Discussion

### A-site magnetically co-doped $Pb_{0.98}Gd_{0.02}(Mg_{1/3}Nb_{2/3})_{0.995}O_3$ relaxor ferroelectric

Temperature dependent ZFC and FC magnetization measurements under 100Oe field are shown in fig. 1(a). ZFC-FC curves show bifurcation below 270K, indicating magnetic interactions introduced with $Gd^{3+}$-doping in the system. The *M*(*H*) isotherm at 5K (fig. 1(b)) shows non-linear magnetization but with no coercive field, indicating absence of long range magnetic ordering.

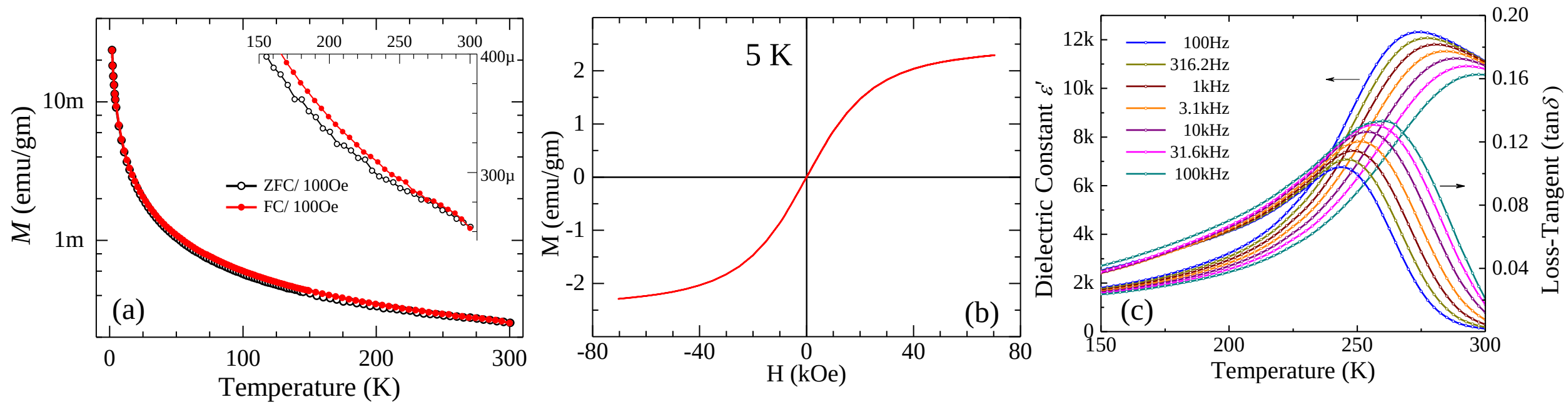

**Figure 1.** (a) Temperature dependence of ZFC and FC magnetization under 100Oe field. (b) Field dependent magnetic isotherm at 5K. (c) Temperature dependence of dielectric constant (left pane) and loss-tangent (right pane) at selected frequencies.

Dielectric constant $\varepsilon'_{\omega}(T)$ have been measured for 2% Gd-doped PMN (B-site vacant) for different frequencies (over 100 Hz to 100 kHz), as shown in fig.1(c) left-pane. The spectra depict frequency dispersive broad maxima ($\varepsilon'_m$ at $T_m$). This shows presence of dipolar regions relaxing at different time scales, in inverse proportionality to their sizes. This thermally-activated relaxation feature also shows up in the observed temperature dependent loss-tangent ($\tan\delta = \varepsilon''/\varepsilon'$) shown in fig.1(c)-right-pane. Low values of $\tan\delta$ (< 1) confirm the intrinsic nature of relaxations, in agreement with the previous report. Since magnetic doping has been performed here, which significantly affects the relaxation features of the system compared to the pure PMN ceramic, the dielectric properties have been further studied under the effect of 9T magnetic field. Figure 2(b)-inset shows the dielectric constant at selected frequencies as a function of temperature under zero and 9T magnetic field. Permittivity increases with the application of *H*-field and the system shows a significantly high magneto-dielectricity of 15% at 225K, as shown in fig. 2(b) main panel. The analyzed frequency dispersion is shown in fig. 2(a), fitted as function of the tan$\delta$-peak temperature, with the following expression;

$$f_p = f_0\left(\frac{T}{T_g} - 1\right)^{z\nu} \tag{3}$$

This relation emulates the spin-glass model, accompanying the absence of long range order in the system. Here, $T_g$ signifies the temperature at which the dynamics is critically slowed down ($\tau \xrightarrow[T \to T_g]{} \infty$). We obtain $T_g$ =211K, approach frequency $f_0$ =54 THz, and exponent $z\nu$ = 16. Here, we emphasize that a Vogel-Fulcher force-fit yields unphysical approach frequency $f_0$, as also previously reported [9]; hence the VFT-model does not describe the dynamics of the system realistically.

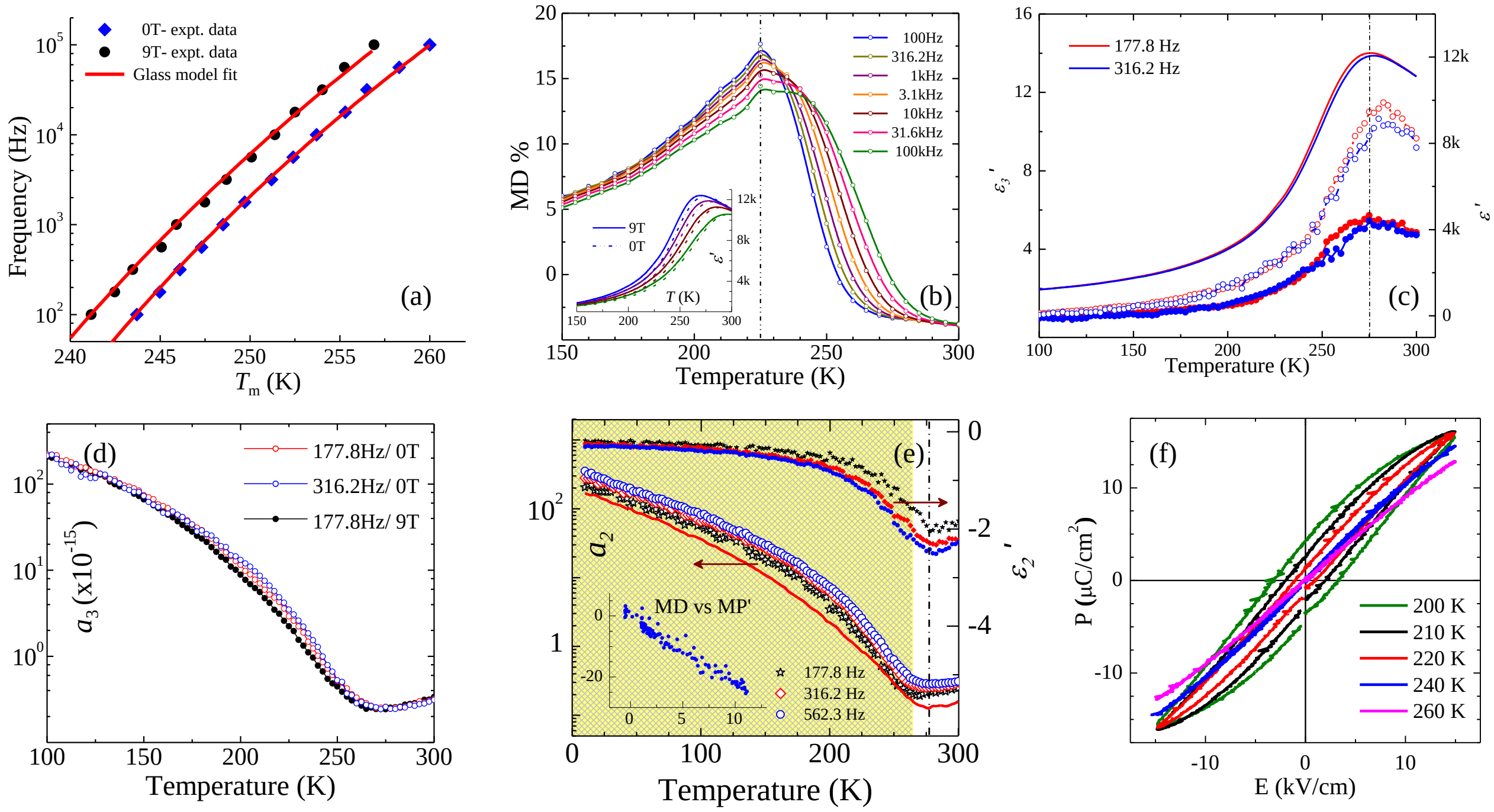

Figure 2 (a) Glassy power-law fit under zero and 9T magnetic field. Symbols are the peak-temperature of tan$\delta$(*f*); solid line is the model fit. (b) Magneto-dielectricity as a function of temperature. Inset: $\varepsilon'(T)$ without and under 9T-field application at selected frequencies. (c) Temperature dependence of second-harmonic signal (left pane) for PMN (open circle) and 2% Gd-doped PMN (solid symbol), with the dielectric constant (in line) for doped-PMN (right pane). (d) Scaled non-linearity $a_3(T)$ for the Gd-doped PGMN. (e) Left-pane: $a_2(T)$, demonstrating an average polarization-rise in the system upon cooling. Right pane: Temperature dependence of the first-harmonic signal. Inset: Anti-regressive changes in MP′ ($P' = -\chi'_2/\chi'^{\,3}$) and MD (vs. implied temperature change). (f) Electric-field induced polarization in the system at different temperatures.

With the application of 9T field $T_m$ (loss-tangent peak temperature) decreases, evidently due to the isothermal decrease in the response time. This indicates an equivalent size-reduction/decorrelation of the dipolar clusters, as also expected from the positive magneto-dielectricity. The frequency-dispersion under 9T-field is also shown in fig. 2(a). It should be noted that the $H$-field application amounts to a devitrification; $T_g$ decreases to 204K, and $f_0$ increases to 82 THz. In the previous reports, similar changes have been observed with the increase in the doping level of Gd in PMN [10]. Here, it can be inferred from our experimental results that the application of magnetic field on the system is analogous to increasing the Gd-doping.

Here, the glass model fit implies the presence of random interactions in the system (with finite averaged squared polarization; $\langle P^2 \rangle \neq 0$). But no definite statements can be made regarding the Nb-rich region-induced non-zero polarization and charge-ordered regions-induced random electric-fields. Hence, to classify the relaxation character in the Gd-doped PMN, non-linear dielectric response that is harmonics' measurements have been performed at low frequencies (instrument-limited). Second harmonic $\varepsilon_3'$ as a function of temperature is shown in fig. 2(c) at the mentioned frequencies. From the previously mentioned expression in eq. 2, it can be said that positive $\varepsilon_3'$ is a consequence of non-zero $\langle P^2 \rangle$ polarization-term (being adequately high than the other $\chi'$-dependent terms in the expression). Here, the decrease in $\varepsilon_3'$ with Gd-doping vis-à-vis the pure PMN confirms that the doped system is driven to glassy state with a reduced squared-polarization; hence the random interactions are prohibited by the Gd-doping. As shown in fig. 2(c) left-pane, $\varepsilon_3'(T)$ follows the dispersion obtained from the fundamental response $\varepsilon'(T)$ (right-pane), reflecting the dipolar dynamics rather well.

Scaled non-linearity parameter $a_3 = \chi'_3/\chi'^4$ [27] is presented in fig. 2(d). The $a_3$-parameter is expected to have a divergent feature for dipolar glasses with vanishing local quenching fields, which is not observed here. $a_3(T)$ for Gd-PMN specimen shows decrease upon cooling, indicating paraelectric-like relaxor phase down to 270K in the system [29]. On cooling further $a_3$ start rising, associated with a transition to glass-like phase. On approaching $T_f$ (210K, obtained from the glass model), rise in $a_3$ strengthens, with frequency dispersion in the vicinity of the respective temperature. The rise in $a_3(T)$, associated with glass-like phase, shifts down to 264K under the application of 9T magnetic field. This indicates magnetically-induced reduction in the dipolar-correlations within the polar clusters, consistent with the positive MD and reduced $T_g$(9T).

First harmonic signal $\varepsilon_2'(T)$ observed as a function of temperature at several frequencies is shown in fig. 2(e)-right-pane. From the expression in eq. 1, it gives a direct evidence of 'polarization' in the system. To get insight about the induced net polarization, we have plotted the scaled $a_2 = -\chi'_2/3B\chi'^3_1$ ($\propto P$) vs. temperature in fig. 2(e) left-pane for PMN and Gd-PMN. Rise in $a_2(T)$ below 270K for Gd-PMN captures an averaged non-zero polarization here, where in the case of pure PMN, $a_2$-signal shows the emergent net polarization below 278K. Here, it should be noted that the polarization doesn't vanishes at $T_f$ (which is expected for the dipolar-glasses), in consistency with the long tail-like feature above $T_f$, observed previously in the PMN system [30]. Upon cooling below $T_f$, unsaturated polarization down to lower temperatures shows that relaxations are mobile even below $T_f$ [30]. For Gd-PMN, effect of the applied magnetic field on $a_2$ is shown in fig. 2(e) inset; consistently, the maximum (-ve) magneto-polarization is found at 225K, concurrent with the maximum MD, shown in fig. 2(b). To check for the actual polarization in the system, $P(E)$ traces measured below 270K are shown in fig. 2(f). Clearly-split loss-free non-linear *P-E* hysteresis loops are witnessed below 240K. The non-linear *P-E* dependence above 240K confirms the correlations amongst the dipoles in the system.

## A site Ca-doped B-site magnetic $La_{0.95}Ca_{0.05}CoO_3$ spin-state perovskite

**Figure 3(a)** presents temperature dependent magnetic susceptibility of $La_{0.95}Ca_{0.05}CoO_3$. System follows Curie-Weiss behavior above 200K;

$$\chi = C/(T-\vartheta_{C\text{-}W}) \qquad (4)$$

Here, Curie-Weiss temperature $\vartheta_{C\text{-}W}$ is obtained as -77.4K, indicating the presence of antiferromagnetic interactions at low temperatures. The Curie constant is obtained as $C = 6.99\times10^{-3}$ emu-K/gm-Oe and hence the moment $\mu(Co^{3+})$ is calculated to be $1.82\mu_B$. Bifurcation in the FC/ZFC curves shows some magnetic correlations below 90K. The downfall in ZFC curve below 10K concurs with the freezing temperature of the dipolar degrees of freedom affiliated to the LS state of the system, as seen later.

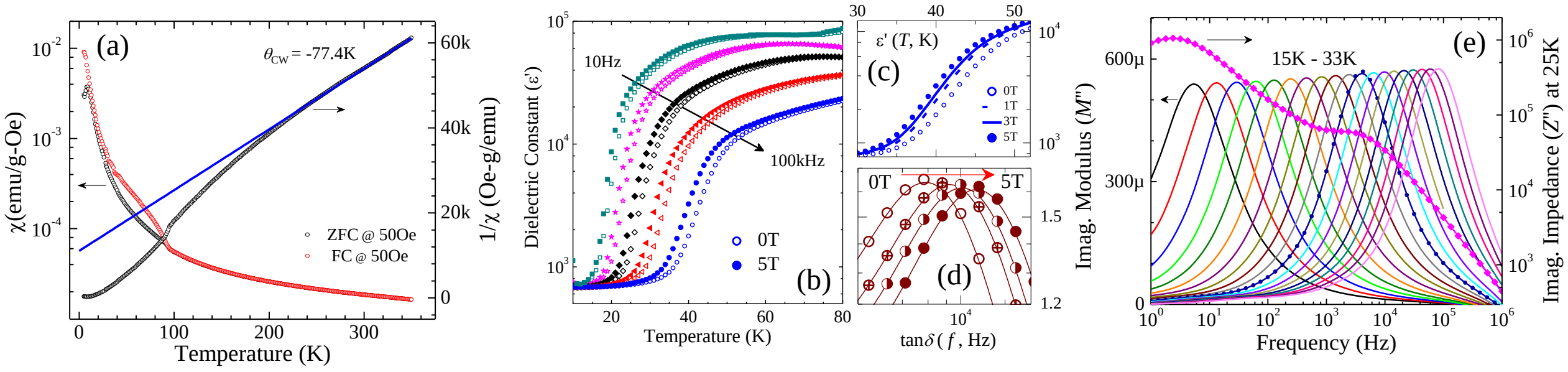

Figure 3. (a) Temperature dependence of ZFC & FC magnetic susceptibility (left pane) and ZFC-inverse magnetic susceptibility (right pane) with Curie-Weiss fit (solid line). (b) Thermal characterization of $\varepsilon'$ under 0T and 5T is shown at several frequencies (100 Hz to 100 kHz). (c) $\varepsilon'(T)$ at 100 kHz under various magnetic field strengths: 0T, 1T, 3T, and 5T. (d) Inset: 30K-tan$\delta(f)$ spectra depicting effect of increasing magnetic field as 0T, 1T, 3T, and 5T. (e) Left pane: isotherms of $M''(f)$ from 15K to 33K at 1K-interval. Right pane: spectral isotherms $Z''(f)$ at 25K.

Complex dielectric constant i.e., $\varepsilon = \varepsilon' + i\varepsilon''$ measured for the sample from 10K to 80K at different frequencies in the range of 1Hz to 100 kHz is shown in **fig. 3(b)-(e).** $\varepsilon'(T)$ in **fig. 3(b)** depict dispersive frequency-dependent steps, with relaxation peaks in the loss-tangent ($\tan\delta = \varepsilon''/\varepsilon'$) and imaginary dielectric modulus ($M'' = \varepsilon''/(\varepsilon'^2 + \varepsilon''^2)$), are shown in **fig. 3(d), (e)**. Here, we emphasize that these kind of anomalies in fundamental dielectric properties {$\varepsilon'(T)$ and $\tan\delta(f)$} have been suggested to indicate multiple phases with different electrical properties [31]. $Co^{3+}$ in the LS state at low temperatures excites to the Jahn-Teller active IS state upon warming, yielding mixed LS-IS state over a broad temperature window, manifesting in a relaxor-like feature. The Jahn-Teller effect with the change in bond-length(s) of the $CoO_6$ octahedra enhances the dielectric constant (>$10^4$) upon warming [31]. To characterize the relaxation processes, impedance and modulus spectroscopy is a powerful tool. Here, we analyze the frequency plots of imaginary modulus ($M''$) and impedance ($Z''$). In **fig. 3(e)**-right pane $Z''(f)$ shows two relaxations, whereas the higher-frequency relaxation is also observed in $M''(f)$ across the overlapping frequency-band. In the $Z''$-spectra, the more-resistive relaxation dominates, whereas in the $M''$-spectra, the less-capacitive one dominates. This confirms that the higher-frequency relaxation peaks observed in the modulus spectra in **fig. 3(e)**-left-pane are the intra-grain contribution. The $f_p(T)$-dispersion is analyzed here, showing a changeover from the high-temperature Arrhenic activation to low-temperature glassy dynamics below 28K, which will be discussed later. Here, the fraction of LS and IS states is tuned by the application of magnetic field. With increase in the applied field strength, dielectric constant $\varepsilon'_f(T, H)$-steps in **fig. 3(c)** depict temperature-downshift and loss-tangent tan$\delta_T(f, H)$-peaks in **fig. 3(d)** undergo frequency-upshift. This evidences that the application of magnetic field is analogous to thermal warming. The aim of the present study is to characterize the relaxor-like dipolar state in greater detail, via study of the harmonic-dielectric response. Furthermore, conductivity and hyper-polarizations (local) coupled to the LS- and IS- magnetic phases are evaluated and analyzed in detail.

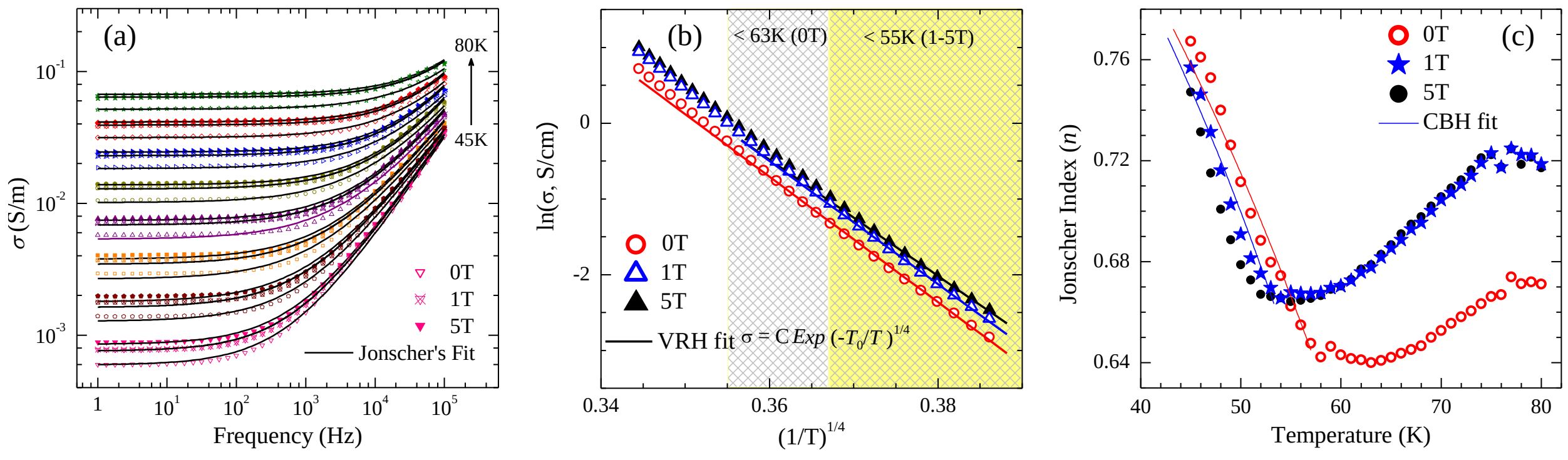

Figure 4. (a) Conductivity measured as a function of frequency (1Hz to 100 kHz) with Jonscher fits (lines) under 0T, 1T, and 5T. (b) $\ln(\sigma)$ vs. $(1/T)^{1/4}$ under 0T, 1T, and 5T-fields, with the VRH-fit (solid line). (b) Jonscher exponent $n(T)$ under 0T, 1T, and 5T.

Owing to the different occupancy of electrons in the LS $\left(t_{2g}^{6}\right)$ and IS $\left(t_{2g}^{5}e_{g}^{1}\right)$ states, conductivity behavior of the system is expected to show the SST-coupled anomaly. **Figure 4(a)** shows the Jonscher power-law fitted ac-conductivity, under zero and 5T magnetic fields. From the Jonscher fits [32], dc-conductivity ($\sigma_{dc}$) and the Jonscher exponent ($n$) are obtained, using the following empirical expression:

$$\sigma(\omega) = \sigma_{dc} + A\omega^{n} \quad (5)$$

At low temperatures, dc-conductivity is found to evidence Mott's variable range hopping (VRH) [33].

$$\sigma_{dc} = C\ Exp\left\{-\left(\frac{T_0}{T}\right)^{1/4}\right\} \quad (6)$$

The VRH-fit is shown in **fig. 4(b),** where upon warming above 65K, $\sigma_{dc}$ of the system shows deviation from the VRH behaviour. In concurrence to the previous report [34], depicting VRH in $LaCoO_3$ with LSS of $Co^{3+}$ ion, this indicates the remarkable presence of LSS of $Co^{3+}$ in $La_{0.95}Ca_{0.05}CoO_3$ up to 65K. The $T_0$ parameter is found $4.6\times10^{7}$ K (slightly reduced, compared to $T_0 = 7.3\times10^{7}$ K for $LaCoO_3$ [34]). From $T_0$, we can also determine the density of electronic states ($N_e$), effectively contributing to the hopping, and the hopping length ($r_{\max}$), using the following expressions [34];

$$T_0 = \left(\frac{21.2\alpha^3}{k_B N_e}\right)\ ;\ r_{\max} = 1.073\left(\frac{1}{\alpha k_B N_e T}\right)^{1/4} \quad (7)$$

Here, $1/\alpha$ is the localization length, which has been taken to be 0.6Å (considering the ionic model for $Co^{3+}$), to estimate the $N_e$ and $r_{\max}$ values. We have also carried out calculations using the same, which is prone to errors from the underestimated localization length (owing to the negligence of covalent bonding). The density of electronic states is obtained as $2.5\times10^{22}$/eV cm$^3$, in agreement with that for the $LaCoO_3$-based semiconductors.

On comparing with $N_e$ ~$1.5\times10^{22}$/eV cm$^3$ ($\pm5\times10^{20}$) for $LaCoO_3$ [34], $La_{0.95}Ca_{0.05}CoO_3$ shows slight increase in the density of electrons participating in the hopping. The maximum hopping length calculated at 50K is ~9.3Å. The decreased value of $r_{\max}$ for $La_{0.95}Ca_{0.05}CoO_3$ compared to that for $LaCoO_3$ (10.4 Å at 50K) [34], is in agreement with the increase in conductivity with Ca-doping, reported in literature [35]. With the application of magnetic field, VRH is observed up to 55K only. $T_0$ decreases to $4.3\times10^7$ K under 1T and further to $3.4\times10^7$ K under 5T. $N_e$ increases to $2.6\times10^{22}$/eV cm$^3$ and $3.3\times10^{22}$/eV cm$^3$ under 1T and 5T respectively. This indicates increase in conduction with the application of magnetic field. In concurrence, $r_{\max}$ decreases as 9.1 Å (50K) and 8.6 Å (50K) under 1T and 5T respectively.

The ac-conductivity power-law index $n(T)$, obtained from the Jonscher fits, is shown in **fig. 4(c)**, depicting a qualitative change in its temperature dependence across 63K. Negative slope of $n(T)$ curve in the range of 45K to 63K indicates correlated barrier hopping (CBH) [36]. Here $n(T)$ can be expressed as;

$$n = 1 - \left( \frac{6k_{\mathrm{B}}T}{W - k_{\mathrm{B}}T\ln(1/\omega\tau_0)} \right) \tag{8}$$

This estimates the hopping energy barrier ($W_b$) to be 24 meV. On warming, $n(T)$ switches to positive slope, depicting Jahn-Teller distortion-assisted small polaron conduction (SPC) above 63K [36]. With the application of magnetic field, the start of spin state transition shifts to lower temperatures ($55\mathrm{K}|_{1\mathrm{T}}$, $53\mathrm{K}|_{5\mathrm{T}} < 63\mathrm{K}|_{0\mathrm{T}}$), as the magnetic field boosts excitation to the ISS. In the following, magneto-conductive features are observed under 1T and 5T magnetic field, where the small polaron conduction is observed down to 56K, with cross-over to CBH upon cooling.

**Figure 5(a)** shows the evaluated MD under 1T. Further, ML is plotted as a function of MD in **fig. 5(b),** where the intrinsic MD-response region is highlighted using the Catalan's criteria [37]. It can be said that as the frequency decreases, the intrinsic MD-response region (i.e., negative slope of positive MD with respect to positive ML) in the system shifts to lower temperatures, as shown in **fig. 5(b)**-inset. Further investigations have been performed via non-linear/harmonic dielectric signals $\varepsilon_2'$ and $\varepsilon_3'$. In **fig 5(c)** presenting $\varepsilon_2'(T)$ for $La_{0.95}Ca_{0.05}CoO_3$, a clear negative-peak anomaly at ~65K is observed, where the signal-strength decreases with the increasing frequency. Similarly, positive-peak signal is observed in the second harmonic signal ($\varepsilon_3'$, **fig.**

**5(d)**) around 70K, whose signal-strength decreases with the increasing frequency. It is noted that no anomaly is observed in the dielectric constant ($\varepsilon'$) across $T_{SST}$-start around 65K. Evidently, the relatively-larger polarizability in the IS-rich phase here is little affected by the much smaller ($\varepsilon'$-reducing) contribution from that of the just-nucleating LS-phase, and thus escapes anomaly-manifestation in the measured fundamental signals. IS and LS states of $Co^{3+}$ have differently-coupled $CoO_6$-octahedral polarization states; $\varepsilon'_{IS} >> \varepsilon'_{LS}$. Nonetheless, the co-emergent interface between IS-matrix and LS-nuclei (each with dipoles exclusively-affiliated to IS or LS states), hosting two very different dipolar entities, is well-formed at $T_{SST}$-start. This heterogeneous dipolar-constitution features non-linear response as the hyper-polarizations ($P$ being proportional to $E^n$, $n$ being positive integer >1), which manifest as anomalies exclusive to the harmonic-signals around $T_{SST}$-start. Non-dispersive nature of the peak-anomalies here in $\varepsilon'_{2,3}$-signals indicates a rather uniform thickness of the IS-LS interface. Associated modulations of interfacial-dipoles' orientation may even configure into multipole (quadrupole/octopole) moments. The negative $\varepsilon'_2$-signal concurrent with the positive $\varepsilon'_3$-signal near 65-70K shows major contribution from the polarization-term (eqs. 1 and 2), which is effected by the Jahn-Teller distortion-assisted changes in the Co-O bond-length(s) of $CoO_6$ octahedra. Under the application of magnetic field, changes in $\varepsilon'_{2,3}(T)$-signals are shown in **fig. 5(c)**-inset and **fig. 5(d)**-inset (bottom-right) respectively. With increase in the applied field, both these anomalies downshift in temperature. This is consistent with the expected in-field extension of the Jahn-Teller distorted ISS to lower temperatures, as also revealed from the field-dependence of VRH-$\sigma_{dc}$ and of the Jonscher-exponent $n$.

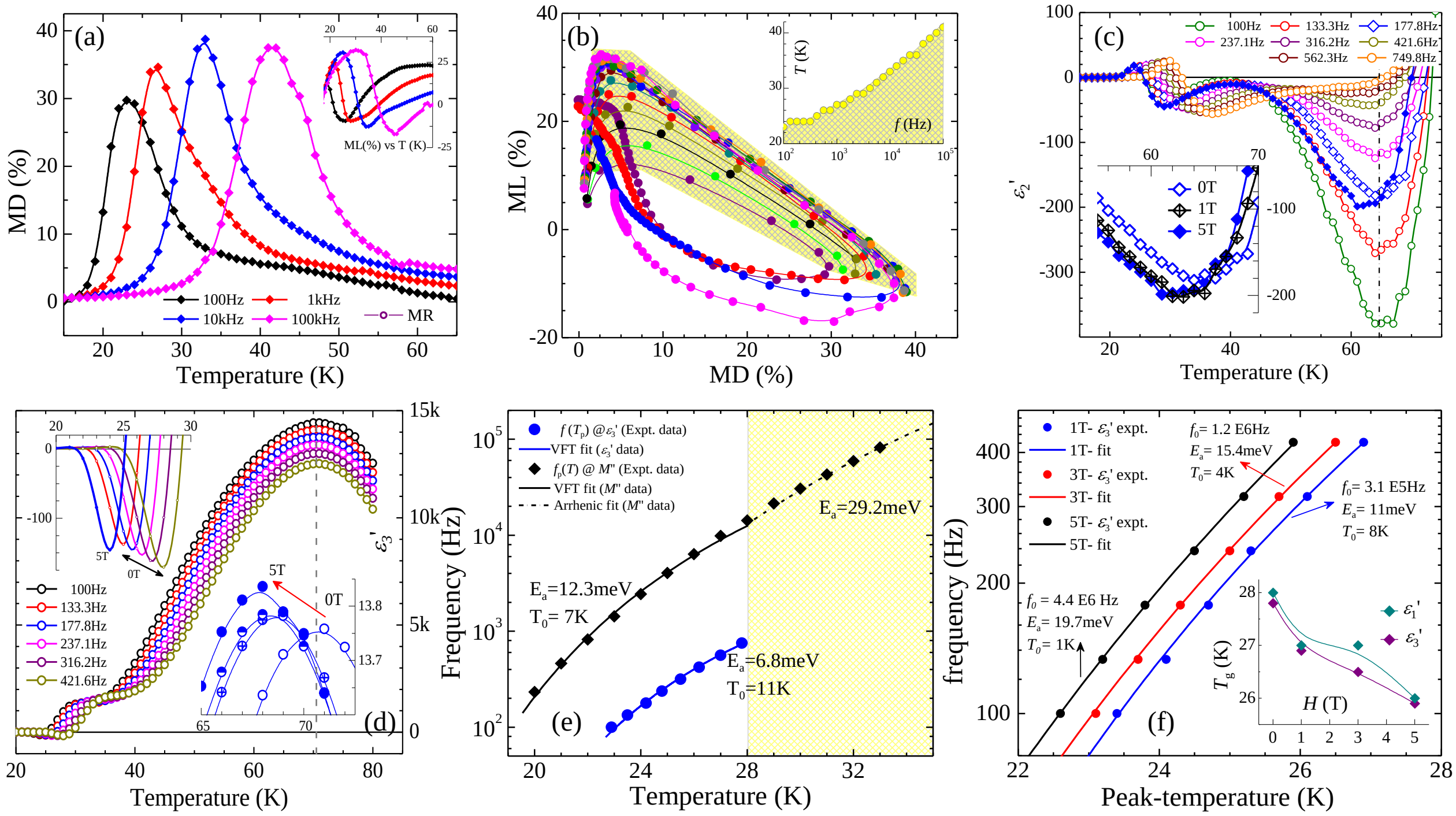

Figure 5. (a) MD($T$) at mentioned frequencies. Inset: Thermal evolution of ML. (b) ML vs. MD at various frequencies, where intrinsic effect is highlighted and corresponding $f(T)$ region is shown in inset. (c) First dielectric-harmonic as a function of temperature in frequency range 100 Hz to 750 Hz. Inset: $\boldsymbol{\varepsilon'_2}(T)$/177.8 Hz under magnetic fields of 0T (open symbols), 1T (crossed symbols), and 5T (solid symbols). (d) Temperature dependence of second dielectric-harmonic with low temperature zoomed-in view (top-left inset). Bottom-right inset: $\boldsymbol{\varepsilon'_3}(T)$/177.8 Hz under increasing magnetic field strength from 0T-1T-3T-5T. (e) frequency-temperature dispersion of relaxations ($M''$ peaks) with VFT-fit (solid line) in fundamental signal, and in second-harmonic. (f) frequency-temperature dispersion with VFT-fit (solid line) for peaks observed in the second-harmonic response under 1T, 3T, and 5T fields. Inset: Magnetic field dependence of $T_g$ obtained from tan$\delta$-peaks (as the crossover from Arrhenius to VFT dispersion) and from $\boldsymbol{\varepsilon'_3}$ (emergence of the lower-$T$ negative peaks).

Here, polarization is a direct consequence of the Jahn-Teller distortion, that is associated with ISS, and $\varepsilon'_{2,3}(T)$-signals well capture every aspect of the harmonic magneto-dielectric effect in the system. Additional, frequency-dispersive anomalies in $\varepsilon'_{3,2}(T)$-signals are observed, upon cooling below 40K and 30K respectively. This corresponds to the region where the bifurcation of ZFC/FC- magnetic susceptibility can be seen to disappear in **fig. 3(a)**. Dispersive nature of these $\varepsilon'_{2,3}(T)$-peaks indicates that the harmonic-signals originate from the (segmented/size-distributed) dipoles affiliated with the LS-state matrix, which coalesces by $T_{\mathrm{SST}}$-end. Upon cooling below 28K, negative-valued peaks in $\varepsilon'_3(T)$ (**fig. 5(d)** top-left inset) evidence relaxor-like short-range dipolar correlations, with zero net macroscopic polarization [27]. In the same regime, $\varepsilon'_2(T)$ shows a changeover from negative- to positive-valued peaks. Across this changeover, the $\varepsilon'_2$-signal strength at lower temperature (positive peaks) decreases to the half of its magnitude at higher temperature (negative peaks), as shown in **fig. 5(c)**. Frequency dispersion $f(T_m)$ for $\varepsilon'_3(T_m)$ follows the Vogel Fulcher timescale-divergence [38];

$$f_p = f_0\, Exp\left(\frac{-E_a}{k_{\mathrm{B}}\left(T - T_0\right)}\right) \tag{9}$$

The approach frequency ($f_0$) is obtained as $9.3\times10^4$ Hz, with activation energy ($E_a$) being 6.8meV and freezing temperature, $T_0$ =11K. Here, $T_0$ concurs with the downfall-changeover in the ZFC-magnetization curve—demonstrating the 'complementary' magneto-electric effect of an electrical event on the magnetic property. **Fig. 5(e)** depicts the fitting profile for $\varepsilon'_3(T_m)$-peaks using the above equation. The VFT- & Arrhenius-fitted $f_m(T)$-dispersion for $M''(f)$ relaxation peaks are also shown. In **fig. 5(e)**, it is clearly observed that the $M''(f)$-relaxation peaks above 28K follow Arrhenic behaviour ($T_0$ =0K); with fitting parameters: $f_0$ =$2.3\times10^9$ Hz and $E_a$ =29.2meV. Below $T_g$ =28K, relaxation peaks depict glassy behaviour with $f_0$ (approach frequency of dipoles) =1.1×10$^7$ Hz, $E_a$ =12.3meV, and $T_0$ =8K. Positive $\varepsilon_2'(T)$ and negative $\varepsilon_3'(T)$ below 28K together indicate dipolar-glass with adequately-low locally polarizable states in the system. Furthermore, the effect of magnetic field on vitreousity can now be depicted much more clearly from the negative peaks in $\varepsilon_3'(T_m)$, as shown in **fig 5(f)**. Temperature-downshift of relaxation-peaks with consecutive decrease in the observed $T_g$'s is observed, upon increase in magnetic field strength. The freezing temperature $T_0$ decreases down to 1K under 5T field application. These decreases signify deterioration of dipolar correlations under applied magnetic field.

## Ferromagnetic double perovskite $La_2NiMnO_6$ multiglass

To examine the ordering of $Mn^{4+}$ and $Ni^{2+}$ ions in $La_2NiMnO_6$ specimen, polarized Raman spectroscopy with parallel polarization (XX) and cross polarization (XY) geometries has been performed. In literature, the B-site cation-ordering with monoclinic symmetry has been confirmed in $La_2CoMnO_6$ with the weakening of the XY-configuration Raman mode intensity at ~650 cm$^{-1}$ vis-à-vis the XX-configuration [39]. **Figure 6(a)** presenting the polarized Raman spectra of $La_2NiMnO_6$ shows that the intense Raman mode (~650 cm$^{-1}$) in the XX-configuration weakens, on changing the configuration to XY-geometry. This confirms the ordered monoclinic structure of the LNMO perovskite. Temperature dependence of magnetization depicts long range ferromagnetic order below 280K, as shown in **fig. 6(b)**. Bifurcation in ZFC and FC upon cooling, with a kink in ZFC is evidence of the antisite disorder in the system. System doesn't show Curie-Weiss behaviour up to 350K. **Figure 6(c)** shows magnetic field dependent *M*(*H*) isotherms at the mentioned temperatures. Magnetic moment of 4.25$\mu_B$/f.u. is observed at 10K under 7T field. This moment value is 0.75$\mu_B$/f.u. less than the theoretical value of 5$\mu_B$/f.u., depicting 7.5% antisite-disorder in the system. The anomalous hump in *M*(*T*) below 50K has been attributed to a glassy state in literature [25].

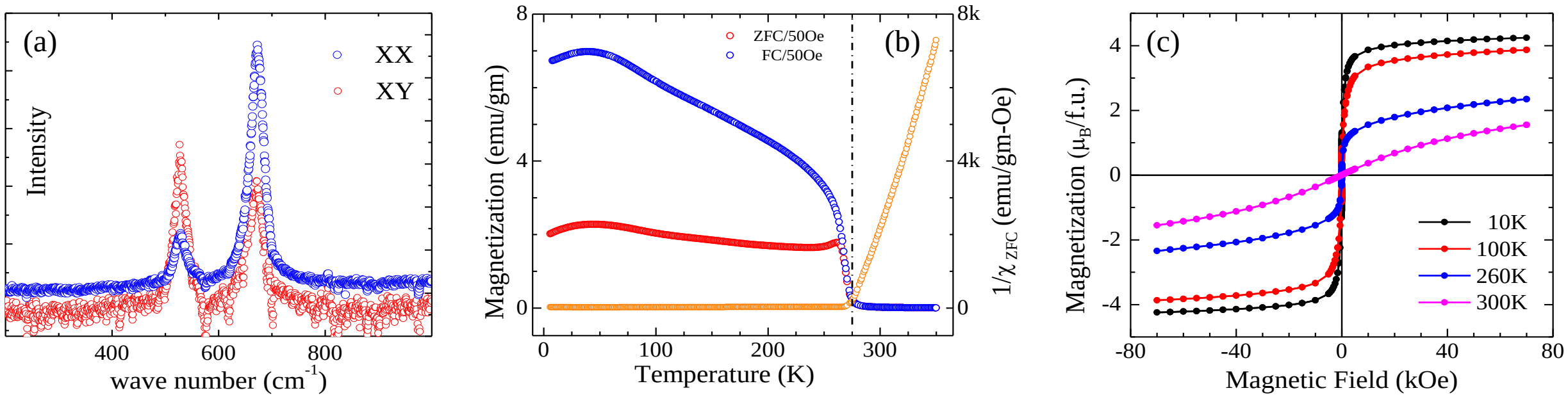

Figure 6.(a) Polarized Raman spectra of $La_2NiMnO_6$. (b) Temperature dependence of ZFC/FC- magnetization (left pane) and inverse susceptibility (right pane). (c) *M*(*H*) isotherms at different temperatures.

Temperature dependence of dielectric constant($\varepsilon'$) depicts orders of magnitude change upon cooling below room temperature down to 10K, as shown in **fig. 7(a)**. Two features can be discerned from $\varepsilon'(T)$ and tan$\delta(T)$; at 10 kHz (say) one relaxation is seen above 250K and another one upon cooling below 200K. At higher temperatures, application of magnetic field tends to decrease the dielectric constant, yielding negative magneto-dielectricity. There is a turn-over to positive magneto-dielectricity upon cooling, as shown in **fig.7(a)-** inset. **Figure 7(c)** presents MD vs. temperature at different frequencies; where the negative MD appears to be associated with the relaxations at higher temperature, with -3.5% maximum value. With the set of relaxations at lower temperatures, positive magneto-dielectricity of ~6% is observed. Loss-tangent ($\varepsilon''/\varepsilon'$) depicts relaxation peaks at lower temperatures, as shown in **fig. 7(b)**. These relaxations have also been reported previously in the literature, featuring thermally-activated Arrhenic dispersion [40]. In literature, no magneto-losses (ML) have been reported, which is important in attribution of genuine magneto-electricity, independent of the magneto-resistive effects.

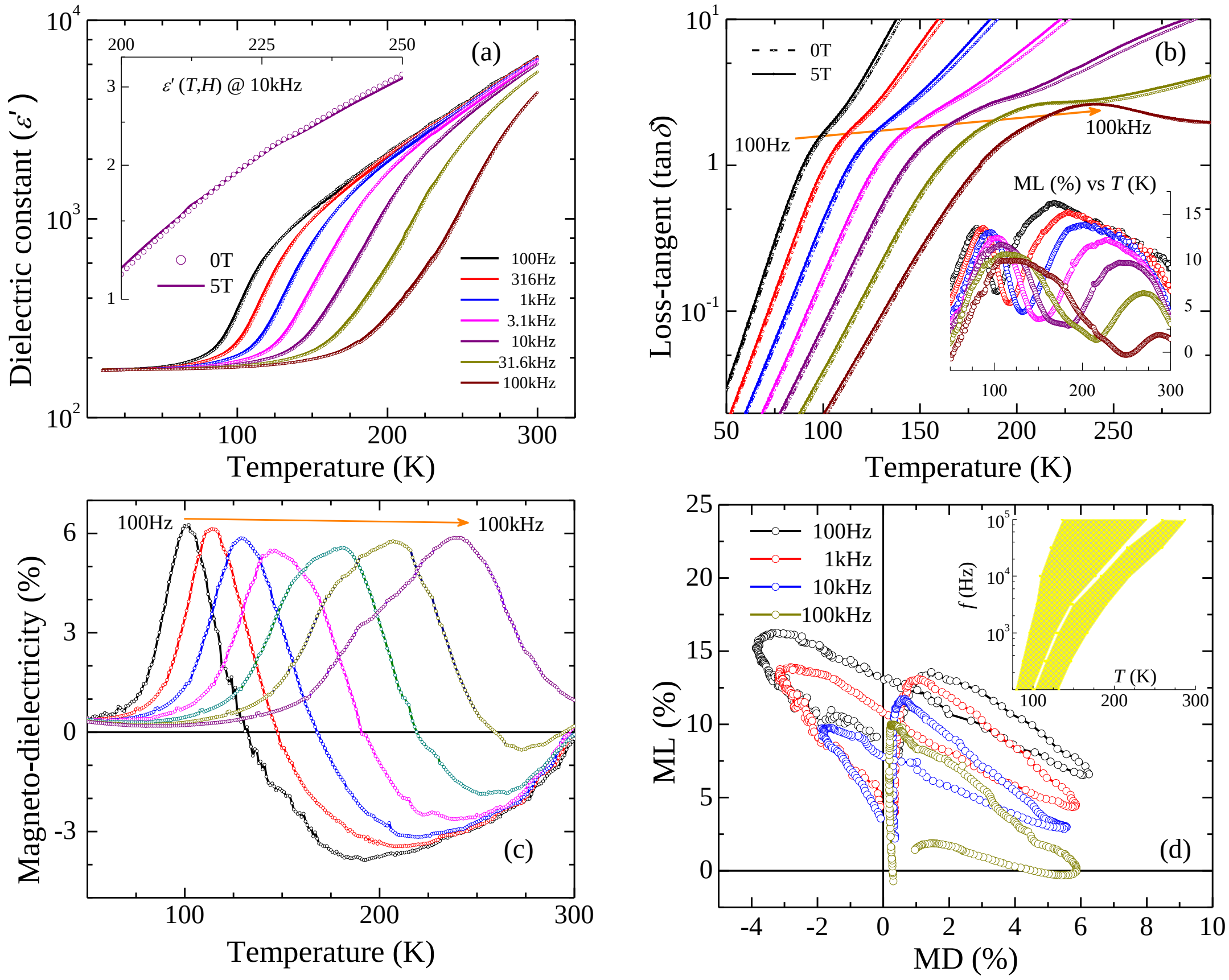

Figure 7(a) Temperature dependent dielectric constant at different frequencies under 0T and 5T field. Inset: zoomed-In view of $\varepsilon'(T)$ at 10 kHz, discerning changes under magnetic field. (b) tan$\delta(T)$ relaxations at different frequencies under 0T and 5T (legends as in (a)). Inset: magneto-losses vs. temperature. (c) Magneto-dielectric effect at different frequencies (legends as in (a)). (d) ML vs. MD. Inset highlights the $f$-$T$ regime where MD is exclusive of the magneto-resistive and Maxwell-Wagner effects.

Here, **fig. 7(b)** inset depicts the temperature dependence of ML, clearly featuring the presence of two different sets of relaxations, of different origin in the system. To obtain the MD-regime exclusive of the extrinsic effects, Catalan's formalism [37] has been utilized. **Figure 7(d)** shows isochrones of ML vs. MD. At high temperatures, negative MD with positive ML is attributed to the interface-dominated MR and Maxwell-Wagner polarization. Corresponding to the intrinsic MD (i.e., positive MD with positive ML), the determined $f(T)$ regions attributed to the genuine ME is presented in **fig. 7(d)**-inset.

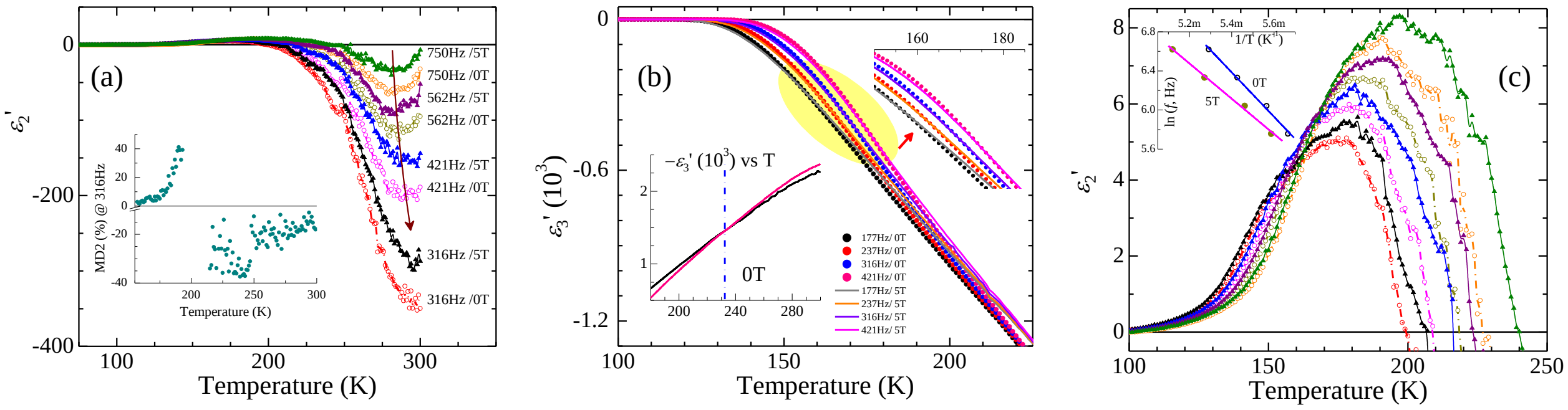

Figure 8(a) Temperature dependence of $\varepsilon_2'$ at different frequencies under 0T and 5T magnetic field. Inset- change in magneto-harmonic effect across 225K. (b) $\varepsilon_3'(T,f)$ under 0T and 5T magnetic field. Left-bottom inset: frequency-crossover in $\varepsilon_3'(T,f)$ under 0T. Righ-top inset: zoomed-in view of the effect of magnetic field on $\varepsilon_3'(T)$. (c) Low-temperature dispersive peak in $\varepsilon_2'(T,f)$ under 0T and 5T (same legend as in (a)). Inset: frequency dispersion of respective peaks with their Arrhenic fits.

In continuum, harmonic dielectric measurements were conducted on the system, which capture the local change in polarization in the system. **Figure 8(a)** shows the temperature dependence of the first-harmonic $\varepsilon_2'$, depicting frequency dependent peaks above 250K, with decrease in signal strength upon increase of the probing frequency. Application of magnetic field tends to further decrease the signal, with negative magneto-harmonic effect at high temperatures. These decreases can be traced to the polarization effect from the free-charges accumulated across the grain-boundaries. With the application of magnetic field, hopping is facilitated in the present ferromagnetic specimen. Decrease in the signal strength of the first-harmonic at high temperatures, both with the frequency-increase and with the application of magnetic field, corroborates with the de-localization effect on the interfacial charges. In this temperature regime, negative $\varepsilon_3'$-signal shown in **fig. 8(b)** too directly excludes the presence of permanent dipoles, as also expected from eq. 2. Upon cooling below 250K, $\varepsilon_2'$ shows positive dispersive peaks, as in **fig. 8(c)**. In sync to this $\varepsilon_3'(f,T)$ shows cross-overs in signal on cooling below 250K, as in **fig. 8(b)**-inset (left-bottom).

At lower temperatures, positive magneto-harmonic effects in both $\varepsilon_2'$ (**fig. 8(a)**-inset) and $\varepsilon_3'$ (**fig. 8(b)**-right-top inset) are observed, tending to zero on cooling down to 100K. Under the effect of magnetic field, the positive peaks in $\varepsilon_2'(T)$ get shifted to higher temperatures, as shown in **fig. 8(c)**. The peaks follow thermally-activated Arrhenic behaviour, with activation energy; $E_a$(0T) =0.195eV and attempt frequency $f_0$(0T) =1.3×10$^8$ Hz. Under 5T-field, the activation energy decreases to $E_a$(5T) =0.155 eV. These activation energies are close to those required for the electron transfer/hopping between $Ni^{2+}$ and $Mn^{4+}$, consistent with the previous reports, based on the fundamental response [25,26,41]. Facilitation of hopping with the application of magnetic field is reflected in the positive magneto-harmonic effects ($MH_{1,2} > 0$) and decreased activation energy ($E_a$(5T) < $E_a$(0T)). Here, the harmonics clearly differentiate and demonstrate the intrinsic hopping-polarization effect at lower temperatures and the interfacial-polarization effect at higher temperatures, a circumstance not resolved by the fundamental signals alone, reported in the literature [25, 26, 40, 41].

## Conclusions

In $Pb_{0.98}Gd_{0.02}(Mg_{1/3}Nb_{2/3})_{0.995}O_3$, significantly high magneto-dielectricity of up to 15% is observed close to the room temperature at 225K. First- and second-harmonic dielectric susceptibilities depict random dipolar interactions in the system. Scaled $a_2$-signal here reveals the emergence of net polarization below 278K, with long tail-like feature above $T_f$ for pure PMN. In the case of Gd-PMN, rise in $a_2(T)$ below 270K, with unsaturated polarization down to lower temperatures, reveals relaxations as active even below $T_f$. Scaled $a_3$-parameter with no divergence features a transition from paraelectric-like relaxor phase to dipolar glasses (under local quenching fields) at 270K. $a_3(T,H)$ for Gd-PMN specimen indicates magnetically-induced reduction in the dipolar-correlations, consistent with the positive MD and reduced $T_g$.

In $La_{0.95}Ca_{0.05}CoO_3$, dc-conductivity features variable range hopping (VRH) below 63K-start of IS-to-LS state transition (SST), while the ac-conductivity behaviour changes over from small polaron conduction (SPC) to correlated barrier hopping (CBH). Dielectric-harmonics here exclusively capture the SST-assisted magneto-electric changes. Around $T_{SST}$(start) ~65K, anomalies marking the Jahn-Teller active polarized state (IS) crossing over to J-T inactive state (LS), are observed in the harmonic-signals alone, featuring a net averaged polarization. Below $T_{SST}$(end) ~30K, harmonics affirm the relaxor-like state featuring no net polarization. Both $T_{SST}$(start) and $T_g$ of the vitreous state below $T_{SST}$(end) shift to lower temperatures with the application of magnetic field, which favors the intermediate spin state. Dynamical freezing temperature $T_0$, of correlated dipoles affiliated to the low-spin state, concurs with the downtrend changeover in $M(T)$.

Using polarized Raman spectroscopy, ordered double perovskite structure of $La_2NiMnO_6$ has been confirmed. Magneto-dielectric effects from intrinsic $Mn^{4+} \leftrightarrow Ni^{2+}$ electron-transfer and extrinsic interfacial-charges are unambiguously resolved by the dielectric harmonics. In general, first- and second-harmonics in combination consistently determine the averaged-finite (local/global) or fictitious (extrinsic) character of polarization. As an important outcome, the investigation of dielectric-harmonics turns out to be indispensable for probing particulate (magneto-) electric phenomena.

## Acknowledgments

Thanks are due to S.M. Gupta (RRCAT) for his help in work on Gd-PMN. We thank Mukul Gupta and Mr. Layanta Behera for XRD measurements. R.J. Choudhary is acknowledged for providing the magnetic data and their discussions. Our sincere thanks extend to Suresh Bhardwaj for facilitating the dielectric measurements. Sumesh Rana is greatly acknowledged for help with the polarized Raman measurements and analysis.

**Supplemenatry File S1**

XRD characterization of $La_{0.95}Ca_{0.05}CoO_3$ specimen: (Goodness of Fit: $\chi^2$ =1.85)

Symmetry: Trigonal (R-3c)

Lattice Parameters: *a* =5.445(72) Å, *b* =5.445(72) Å, and *c* =13.10(347) Å

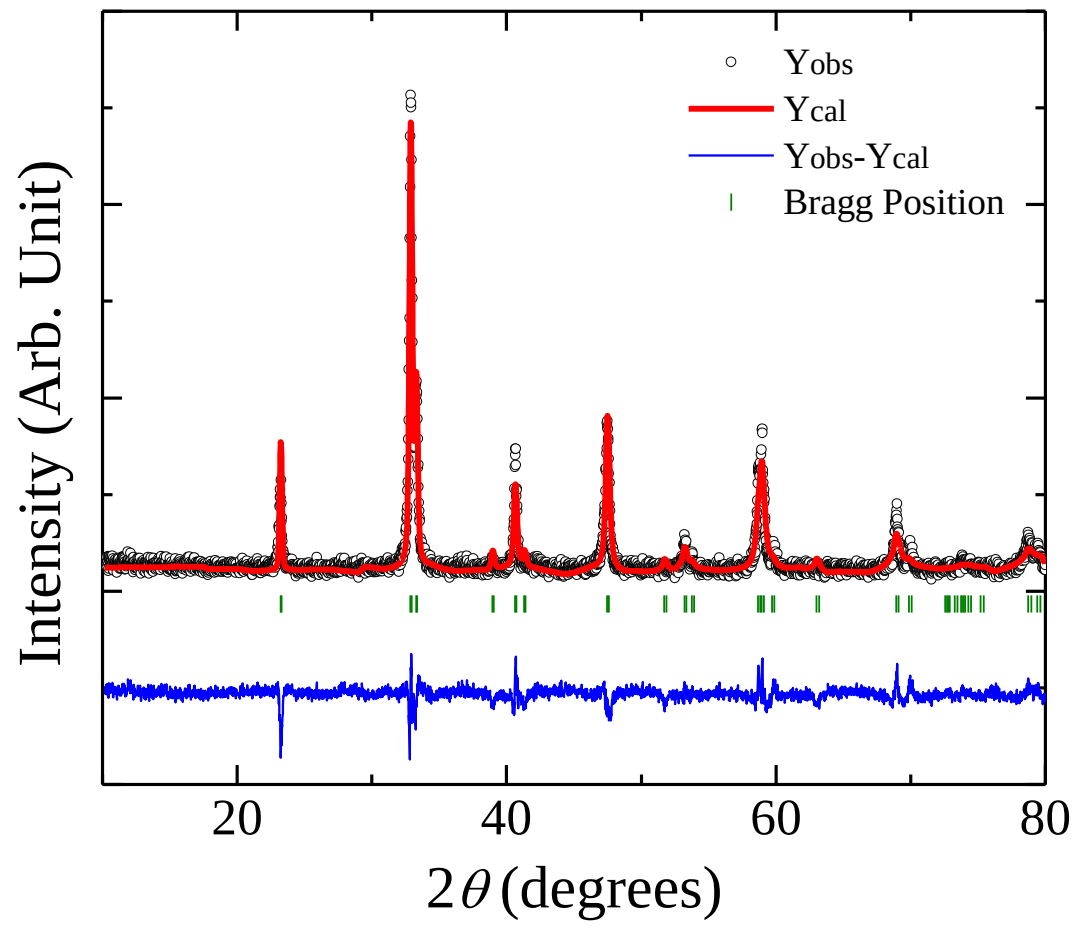

XRD characterization of $La_2NiMnO_6$ ceramic: (Goodness of Fit: $\chi^2$ =2.21)

Symmetry: Monoclinic (P121/n1)

Lattice Parameters: 5.510(04) Å, 5.461(63) Å, 7.739(93) Å, and $\beta$ =90.04718º

Minor Phase (4.11%): NiO

Symmetry: Cubic (Fm3m)

Lattice Parameters: *a* = *b* = *c* =4.178(31)Å

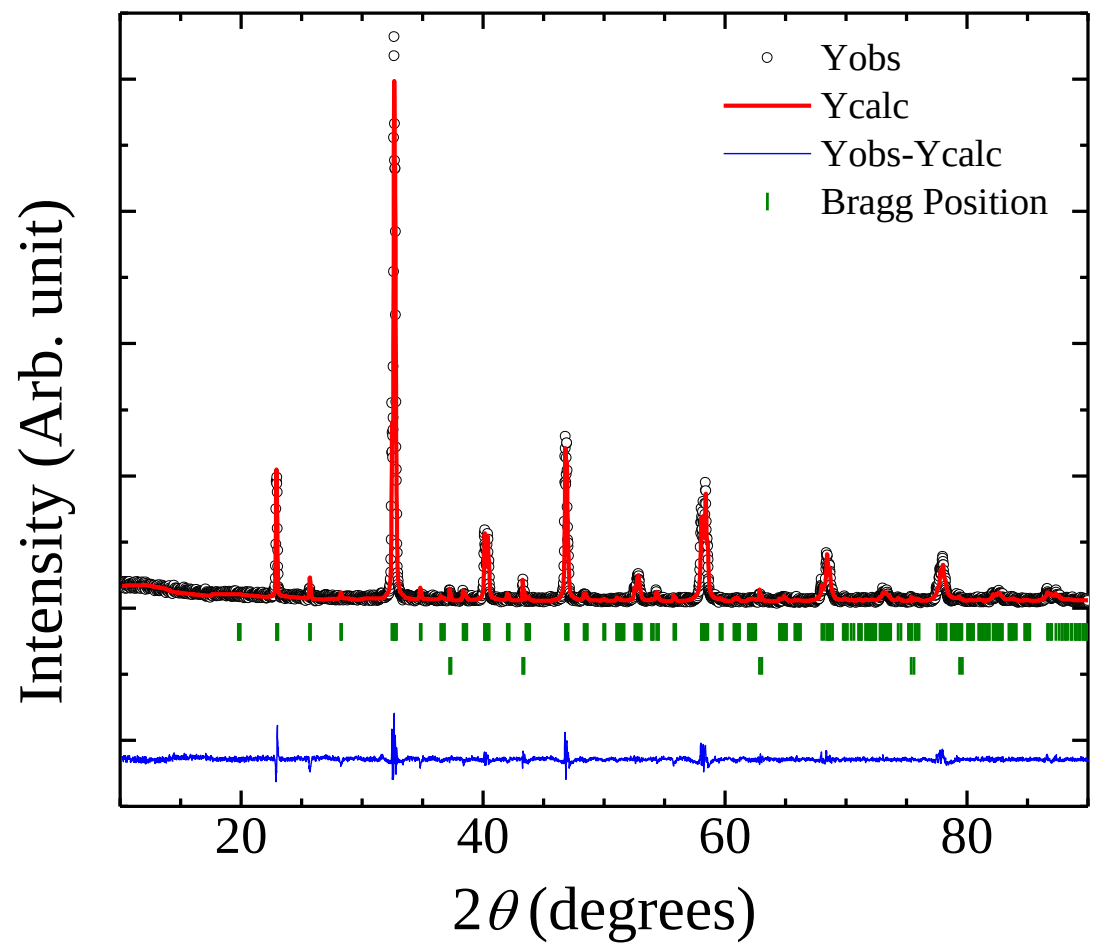